\documentclass[prb,twocolumn,showpacs,aps,superscriptaddress,floatfix]{revtex4-2}
\usepackage{amsmath}
\usepackage{amssymb}
\usepackage{amsfonts}
\usepackage{bm}
\usepackage[dvips]{graphicx}
\usepackage{color}
\usepackage{ulem}
\usepackage[unicode, pdftex]{hyperref}
\usepackage{verbatim}
\begin{document}
\title{Interaction between spin and Abrikosov vortices in doped topological insulators}
\author{A.V. Kapranov}
\affiliation{Dukhov Research Institute of Automatics, Moscow, 127055 Russia}
\affiliation{Moscow Institute of Physics and Technology, Dolgoprudny,
    Moscow Region, 141700 Russia}
\affiliation{Institute for Theoretical and Applied Electrodynamics, Russian
    Academy of Sciences, Moscow, 125412 Russia}

\author{R.S. Akzyanov}
\affiliation{Dukhov Research Institute of Automatics, Moscow, 127055 Russia}
\affiliation{Moscow Institute of Physics and Technology, Dolgoprudny,
    Moscow Region, 141700 Russia}
\affiliation{Institute for Theoretical and Applied Electrodynamics, Russian
    Academy of Sciences, Moscow, 125412 Russia}

\author{A.L. Rakhmanov}
\affiliation{Dukhov Research Institute of Automatics, Moscow, 127055 Russia}
\affiliation{Institute for Theoretical and Applied Electrodynamics, Russian
    Academy of Sciences, Moscow, 125412 Russia}

\begin{abstract}
In the topological superconductor with the nematic superconductivity in the $E_u$ representation, it is possible to have different types of vortices. One is associated with the vorticity in the particle-hole space and corresponds to the Abrikosov vortex. Another type corresponds to the vorticity in the spin space and is called the spin vortex. We study the interaction of the Abrikosov vortex with the spin vortices. We derive the free energy of the sample with the Abrikosov and the strain-induced spin vortices using the Ginzburg-Landau approach for the two-component superconducting order parameter. We calculate the critical strain at which the spin vortex is formed. We show that the spin vortex and the Abrikosov vortex attract to each other, and, as a result, they have a common core. We apply Bogoliubov–de Gennes equations to study electronic states in a combined vortex structure. We show that no zero-energy states (Majorana fermions) are localized near the common vortex core of the Abrikosov vortex and the spin vortex of any type. Possible experimental realization is discussed.
\end{abstract}
\maketitle

\section {Introduction}

In the last years, nematic superconductivity in doped topological insulators has received a lot of experimental evidence~\cite{Yonezawa2019, shen2017, Chen2019, Frohlich2020, Kuntsevich2018, Schmidt2020, Das2020}. In topological insulators such a $\text{A}_\text{x}\text{Bi}_2\text{Se}_3$, where $\text{A}=\text{Cu}$, $\text{Nb}$, $\text{Sr}$, the cooper pairs have a spin triplet pairing ~\cite{Matano2016,Fu2010,Venderbos2016}. The superconducting order parameter is a two-component real-valued vector with $E_u$ symmetry in this materials~\cite{Fu2010}. The nematic superconductivity has a non-trivial coupling with a strain. This coupling leads to interesting phenomena such as a twofold symmetry of the in-plane second critical field~\cite{Venderbos2016, Kuntsevich2018,KuntsevichAkzyanov2018}, spontaneous strain~\cite{Akzyanov2020}, and strain-induced spin vortices~\cite{Akzyanov2021}. 

Superconductivity and superfluidity are rather similar phenomena. In particular, we have a similar picture for topological defects in the topological superconductors with the spin-triplet pairing and in the B phase of superfluid $^3$He~\cite{Volovik_book,Chiu2016}. Along with the usual (mass) vortex, in the B phase of $^3$He, it is possible to create a spin vortex. This topological defect does not change the global phase of a wave function but changes the phase associated with the direction of the spin. Thus, in the spin vortex, we observe a vorticity of the spin~\cite{Volovik_book,Korhonen1993}. In Ref.~\cite{Korhonen1993} it is shown that in the rotating vessel, the mass vortex (analog of the Abrikosov in superconductors) and the spin vortex attract to each other and, consequently, their cores can overlap. It means that a combined vortex can be created and stabilized in the rotating vessel called a spin-mass vortex.

The spin vortices in the context of the superconductivity have been briefly discussed for $\left(p_x+ip_y\right)_\uparrow\left(p_x-ip_y\right)_\downarrow$ topological superconductors~\cite{Chiu2016}. In Ref.~\cite{Akzyanov2021}, it was shown that in the nematic topological superconductors, two possible types of topologically different spin vortices could be realized. In the type I spin vortex, there exist zero-energy states localized near the vortex core. These states can be identified as Majorana-Kramers pairs. In the type II spin vortex, there are no localized zero-energy states.

The Majorana fermion is an excitation with non-Abelian statistics. The study of such non-Abelian states is one of the hot topics in condensed matter physics for different reasons. For example, these exotic excitations could be a basis for topologically protected quantum computations~\cite{Kitaev2003}. The Majorana fermions can be localized on the topological defect in the system. The core of the Abrikosov vortex can be treated as a topological defect~\cite{Maki2006}. The existence of the Abrikosov vortex in the topological superconductors is a well-known fact~\cite{Zhao2018, Nayak2021, Tao2018}, and the Majorana fermions can be localized in the core of the Abrikosov vortex~\cite{Fu2008, Akzyanov2015}. In this context, it is of interest to analyze the interaction between Abrikosov and spin vortices and answer the question of whether the existence of the Majorana fermions is possible in such a two vortices structure.    

Here, we first clarify the Abrikosov vortex's structure in the nematic topological superconductor with vector order parameter using the Ginsburg-Landau (GL) approach. Then, we write down the free energy of the spin vortex and derive the value of the critical strain that generates the spin vortex, either type I or II. Finally, we consider the case of coexisting the Abrikosov and spin vortices and show that they should attract each other. We argue that in the ground state, these two vortices have a common core, that is, the spin vortex can be considered as a pinning center for the Abrikosov vortex. Using the Bogoliubov-de Gennes (BdG) formalism, we study the electronic states in the nematic superconductor with the Abrikosov and spin vortex having the common core. We observe that there are no zero-energy states localized near the common vortex core either in the case of the type I spin vortex or type II one.   

\section{Abrikosov and spin vortices}

In this section, we analyze the properties of the Abrikosov and spin vortices. Two spatial scales characterize the Abrikosov vortex. First is the size of the vortex core, where the superconducting order parameter is suppressed, which is of the order of the superconductor coherence length $\xi$. Second is the magnetic size of the vortex, which is of the order of the London penetration depth $\lambda$. It is known that the considered nematic superconductors are superconductors of the second type~\cite{Zhao2018, Nayak2021,Tao2018}. Hence, we assume that $\xi\ll\lambda$. 
The spin vortex has two characteristic scales, as well~\cite{Akzyanov2021}. First is the core size $\xi_{I}$ and $\xi_{II}$ for the vortex of type I or II, respectively, and in any realistic case, $\xi_{I,II}$ are of the order of $\xi$. The second scale $l_u$ is the size of the region in which the vorticity of the vector order parameter is observed. This value is determined by the applied external force, the sizes of the sample, or structural defects. The value $l_u$ can be either macroscopic or microscopic, depending on the above factors.

\subsection{Ginzburg-Landau functional}

For the reader's convenience, we present here the expression for the GL functional in a general form. {\color{black}This GL functional can be obtained microscopically from the Hamiltonian of the doped topological insulator with the nematic superconductivity, see e.g.~\cite{Khokhlov2021}.} In the case of a homogeneous phase and the absence of the magnetic field, the GL free energy can be written as~\cite{Venderbos2016} 
\begin{equation}
    \!\!F_0\!=\!a\!\left(\!\left|\Delta_1\!\right|^2\!\!+\!\left|\Delta_2\!\right|^2\!\right)\!+\!B_1\!\!\left(\!\left|\Delta_1\!\right|^2\!\!+\!\left|\Delta_2\!\right|^2\!\right)^2\!\!\!+\!B_2\!\left|\Delta^*_1\Delta_2\!-\!\Delta_1\Delta^*_2\right|^2\!,
\end{equation}
where $\vec{\Delta}=\left(\Delta_1,\Delta_2\right)$ is the vector order parameter, $a\propto T-T_c<0$ and $B_{1,2}>0$ are the GL coefficients. The considered topological superconductor has an anisotropic layered structure. We assume that the plane $(x,y)$ coincides with the crystallographic layers, and the magnetic field is directed perpendicular to this plane along the axis $z$. Then, we have the following electromagnetic contribution to the GL free energy
\begin{equation}\label{eq:F_m}
     F_\text{M}=\frac{\left(\nabla\times\mathbf{A}\right)^2}{8\pi}-\frac{\nabla\times\mathbf{A}\cdot\mathbf{H_0}}{4\pi},
\end{equation}
where $\mathbf{H}_0$ is the external magnetic field and $\mathbf{A}$ is the vector potential. 
There is another term associated with the magnetic field $\mathbf{H}=\nabla\times\mathbf{A}$. It is the contribution to the GL free energy related to the transverse Zeeman magnetization in the $E_u$ superconductor~\cite{Venderbos2016,Khokhlov2021}
\begin{equation}
    F_\text{Zeeman}=-2ig_\text{eff}\mu_\text{B}H\left(\Delta_1\Delta^*_2-\Delta^*_1\Delta_2\right),
\end{equation}
where $g_\text{eff}$ is a GL coupling constant between superconductivity and the Zeeman field. According to Ref.~\cite{Khokhlov2021}, we can neglect the magnetization, since the term $F_\text{Zeeman}/F_M\sim T^2_\text{c}/\mu^2\ll1$.

As usual in the GL theory of superconductivity, the electromagnetic field causes the coordinate dependence of the order parameter and couples with it through the gauge-invariant gradients $D_j=-i\hbar\partial_j+(2e/c)A_j$. As a result, we should consider additional gradient terms in the GL free energy, which are allowed by the crystal symmetry. The considered topological superconductor has a hexagonal crystal symmetry, and the corresponding contribution is~\cite{Venderbos2016,Sigrist1991}
\begin{align}\label{eq: gradient_general}
    &F_\text{D}=\nonumber J_1\left(D_i\Delta_a\right)^*D_i\Delta_a+J_2\epsilon_{ij}\epsilon_{ab}\left(D_i\Delta_a\right)^*D_j\Delta_b\\\nonumber
    + & {} J_3\left(D_z\Delta_a\right)^*\!D_z\Delta_a\!+\!J_4\!\!\left[\left|D_x\Delta_1\right|^2+\left|D_y\Delta_2\right|^2-\left|D_x\Delta_2\right|^2\right.\\\nonumber
    & {} \left.-\left|D_y\Delta_1\right|^2+\left(D_x\Delta_1\right)^*D_y\Delta_2+\left(D_y\Delta_1\right)^*D_x\Delta_2\right.\\
    & {} \left.+\left(D_x\Delta_2\right)^*D_y\Delta_1+\left(D_y\Delta_2\right)^*D_x\Delta_1\right],
\end{align}
where summation is implied over repeating indices $i=x,y$, $a=1,2$, $\epsilon_{kl}$ is the Levi-Chivita symbols, $J_{1,2,3,4}$ are phenomenological GL coefficients, and $J_1>J_4$. 

We also assume that the crystal lattice of the sample is deformed by force. The coupling of the strain with the superconductivity gives rise to an additional term in the GL free energy~\cite{Venderbos2016}
\begin{align}\label{eq:Fu}
    F_\text{u}\nonumber&=g_N\left(u_{xx}-u_{yy}\right)\left(\left|\Delta_1\right|^2-\left|\Delta_1\right|^2\right)^2\\
    & {} +2g_Nu_{xy}\left(\Delta^*_1\Delta_2+\Delta_1\Delta^*_2\right),
    \end{align}
where $u_{ik}$ are the components of the strain tensor and $g_N$ is a GL coupling constant between the order parameter and the strain. The total GL free energy is the sum of all listed above terms 
\begin{equation}\label{eq: gl_total_1}
    F_\text{GL}=F_0+F_\text{D}+F_\text{M}+F_\text{u}.
\end{equation}

\subsection{Abrikosov vortex}
\label{Abrikosov_vortex}


The magnetic field related to the isolated Abrikosov vortex is weak and does not affect the value of the order parameter. 
Since $B_2>0$ in $F_0$, the ground state order parameter is a real valued vector in the absence of the magnetic field. Thus, the phase of the order parameter $\theta$ arises only due to the magnetic field produced by the Abrikosov vortex. Far from the core of this vortex, the local value of the order parameter $\Delta=\Delta_0\exp{\left(i\theta\right)}$ is dictated by the structure of the spin vortex. Under assumptions made, the magnetic field distribution in the Abrikosov vortex can be obtained following a standard procedure~\cite{tinkham2004introduction}. The calculation details are presented in Appendix \ref{sec:a}. Here we write down only the results. 

The magnetic field in the vortex away from its core can be written as
\begin{equation}\label{eq:AbV_field}
H(\rho)=\frac{\Phi_0}{2\pi\lambda^2}K_0\left(\frac{\rho}{\lambda}\right),\,\,\,
\rho^2=\frac{x^2}{1+k}+\frac{y^2}{1-k},
\end{equation}
where $K_0(r)$ is the zero-order MacDonald's function, $\Phi_0=\pi\hbar c/e$ is the magnetic flux quantum, $k=J_4/J_1$, $\lambda$ and $\xi$ are the effective London penetration depth and coherence length
\begin{equation}\label{eq:lambdaxi}
\lambda^2=\frac{c^2}{32\pi e^2J_1(1-k^2)\Delta^2_0},\,\,\,
\xi^2=\frac{J_1\hbar^2}{2B_1\Delta^2_0},
\end{equation}
$\kappa=\lambda/\xi\gg 1$ is the GL parameter, and $\Delta^2_0=-\left[a+g_N\left(u_\text{xx}-u_\text{yy}\right)\right]/2B_1$
is an equilibrium order parameter. Note that the lines of the current in the Abrikosov vortex have an elliptic geometry and the expression for the first critical has a usual form for the type II superconductor~\cite{tinkham2004introduction}. 

\subsection{Spin vortex}
\label{sec:sv}

According to Ref.~\cite{Akzyanov2021}, the strain can generate the spin vortices of two types in the nematic superconductor. These vortices have normal cores, like the Abrikosov vortex. The structures of the order parameter in the spin vortices of types I and II are 
\begin{equation}\label{eq:spin_V_structure}    
\!\!\!\!\!\!\!\vec{\Delta}_\text{I}\!\!=\!\!\Delta(r,z)\!\left(\cos{\varphi},\sin{\varphi}\right),
\,\,\vec{\Delta}_\text{II}\!\!=\!\!\Delta(r,z)\!\left(-\sin{\varphi},\cos{\varphi}\right),
\end{equation}
where $(r,\varphi,z)$ are the cylindrical coordinates. {\color{black} The Majorana-Kramers pair exists near the core of the vortex of type I, and it does not exist near the core of the vortex of type II \cite{Akzyanov2021}}. The spin vortex arises when the strain exceeds some critical value. This value depends on the applied force and the boundary conditions of a particular problem. We assume here that the mechanical problem has cylindrical symmetry. In this case, in the cylindrical coordinates, we have $u_{xx}-u_{yy}=u(r,z)\cos{2\varphi}$ and $2u_{xy}=u(r,z)\sin{2\varphi}$. We consider the simplest and the most illustrative case, assuming that the strain amplitude $u(r,z)$ is constant within the cylinder with radius $l_u$ (see Fig.~\ref{FIG:1}). We also neglect the unimportant $z$-dependence of the problem values and assume that the strain is not too large, that is, $|g_Nu|< |a|$. 

First, we calculate the GL free energy associated with creating the spin vortex of type I or II. The process of calculations is described in the Appendix \ref{sec:b}. Here we present only the result 
\begin{align}
    F^\text{I(II)}_\text{SV}=&\nonumber\frac{\pi}{4B_1}\left[g_Nu\left(\mp2a+3g_Nu\right)l^2_u\right.\\
    & {} \left.-4\left(J_1\pm J_4\right)\left(a\mp g_Nu\right)\ln{\frac{l_u}{\xi_\text{I(II)}}}\right].
\end{align}
where $F^\text{I(II)}_\text{SV}$ means the GL free energy of the spin vortex of type I(II) and $\xi_\text{I}=\xi\sqrt{1+k}$ and $\xi_\text{II}=\xi\sqrt{1-k}$ are effective coherence lengths (or core sizes) for the vortex of type I and II, respectively.

The spin vortex arises in the strained nematic superconductor if $F^\text{I(II)}_\text{SV}<0$. After a simple algebra, we obtain that the existence of the spin vortex is thermodynamically favorable if the deformation $u$ lies within the limits
\begin{equation}\label{eq:crit_strain}
    \left|\frac{2}{3}\left(\frac{\xi_\text{I(II)}}{l_u}\right)^2\ln{\frac{l_u}{\xi_\text{I(II)}}}\right|<\left|\frac{g_Nu}{a}\right|\leq\frac{2}{3}.
\end{equation}
The type of the spin vortex depends on the sign of the value $g_Nu$: the vortex is of type I if $g_Nu<0$ and of type II if $g_Nu>0$. We also see from the condition Eq.~\eqref{eq:crit_strain} that the size of the deformed area should be large enough to generate the spin vortex. In particular, the spin vortex does not arise if $l_u\ll\xi$. {\color{black} A characteristic value of the ratio $|g_Nu/a|\approx 0.2\div 0.6$ for the topological superconductors Bi$_2$Se$_3$ was extracted from the experimental data in Refs.~\cite{KuntsevichAkzyanov2018,PhysRevB.104.L220502}. Thus, we can conclude that the conditions~\eqref{eq:crit_strain} can be satisfied in a real experiment.}

\section{Interaction between Abrikosov vortex and spin vortex}\label{sec:interaction}

\begin{figure}[ht]
    \centering
    \includegraphics[width=0.45\textwidth]{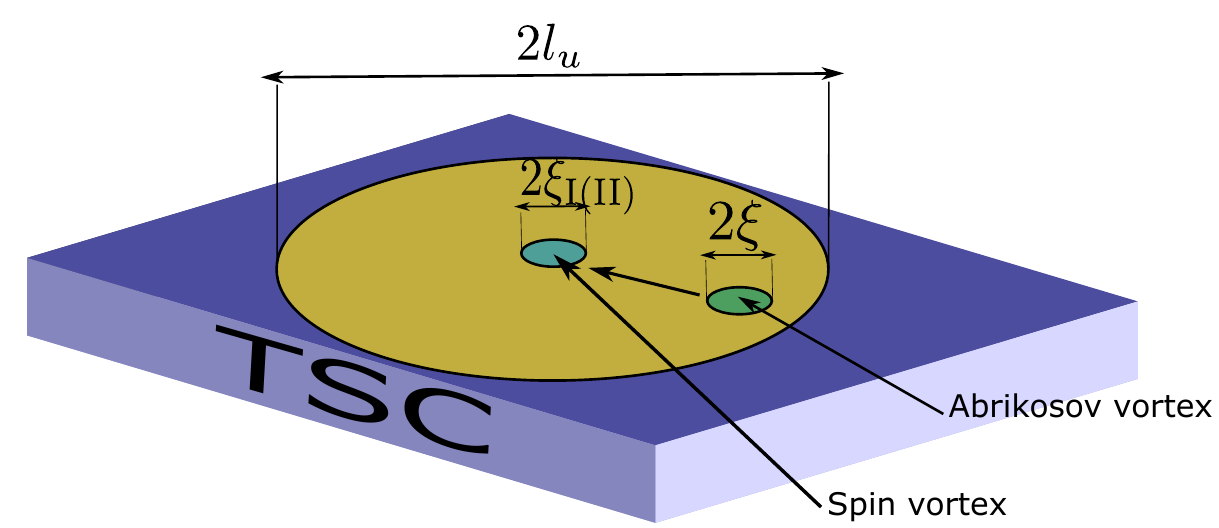}
    \caption{A schematic picture of the system. TSC is the topological superconductor, the deformed region is indicated by yellow, and the Abrikosov and spin vortices are shown by green.}  
    \label{FIG:1}
\end{figure}

Now we consider a strained sample with the Abrikosov and spin vortices in zero external magnetic fields. Let the center of the spin vortex be located at the coordinate origin, and the center of the Abrikosov vortex is at the point $(x_0,y_0)$, see Fig.~\ref{FIG:1}. It is convenient to characterize the distance between vortices by a radius $\rho_0$ in the elliptic coordinates $\rho_0^2=x_0^2/(1+k)+y_0^2/(1-k)$. We assume that $l_u>\rho_0>2\xi$ to vortex
started interacting. If the free energy decreases with the decrease of $\rho_0$, the spin and Abrikosov vortices attract each other. 

The Abrikosov vortex changes the phase of the order parameter. Following a standard approach, we make the following gauge transformation 
\begin{equation}\label{eq:gauge}
    \mathbf{A}=\frac{1}{2}\nabla\chi,\quad\theta=\frac{\pi}{\Phi_0}\chi,
\end{equation}
where $\chi$ is the scalar potential, vector potential $\mathbf{A}=(A_x,A_y,0)$ is determined from Eq.~\eqref{eq:AbV_field} as $\nabla\times\mathbf{A}=\mathbf{H}$, and $\theta$ is the phase of the order parameter. From the gauge transformation Eq.~\eqref{eq:gauge} we obtain the following equations for the scalar potential 
\begin{equation}\label{eq:grad_scalar}
\begin{cases}
    \nabla_x\chi\!=\!\frac{y\Phi_0}{\sqrt{1\!-\!k^2}\pi\lambda\rho}K_1\!\left(\!\frac{1}{\lambda}\!\sqrt{\!\frac{\left(x\!-\!x_0\right)^2}{1+k}\!+\!\frac{\left(y\!-\!y_0\right)^2}{1-k}}\right)\!,\\
    \nabla_y\chi\!=\!\frac{-x\Phi_0}{\sqrt{1\!-\!k^2}\pi\lambda\rho}K_1\!\left(\!\frac{1}{\lambda}\!\sqrt{\!\frac{\left(x\!-\!x_0\right)^2}{1+k}\!+\!\frac{\left(y\!-\!y_0\right)^2}{1-k}}\right)\!,
\end{cases}
\end{equation}
where  $K_1$ is the modified Bessel function. Thus, the order parameter can be presented as 
\begin{align}\label{eq:order}
    \vec{\Delta}_\text{I}&=\Delta(r,z)\exp{\left(i\frac{\pi}{\Phi_0}\chi(r,\varphi)\right)}(\cos{\varphi},\sin{\varphi}),\\\nonumber
    \quad\vec{\Delta}_\text{II}& {}=\Delta(r,z)\exp{\left(i\frac{\pi}{\Phi_0}\chi(r,\varphi)\right)}(-\sin{\varphi},\cos{\varphi}).
\end{align}
We substitute the gradients of the scalar potentials Eqs.~\eqref{eq:grad_scalar} and the order parameters Eqs.~\eqref{eq:order} in the GL free energy Eq.~\eqref{eq: gl_total_1}. As a result, the GL free energy of the considered two vortex system can be presented in the form 
\begin{equation}\label{eq:fsmv}
    F^\text{I(II)}_\text{SMV}\!=\!\!\!\int\!\frac{ a\!\mp\!g_Nu}{4B_1}\left[-a\!\mp\!3g_Nu\!-\!\frac{2J_1\!\left(1\!\pm\!k\right)}{x^2+y^2}\right]\!\!\,dV+F^\text{I(II)}_\text{int}.
\end{equation}
{\color{black} The first term is the free energy associated with the existence of the spin vortex. The contribution in the free energy $F^\text{I(II)}_\text{int}$ includes the terms dependent on the distance between the spin and Abrikosov vortices. (Note, we omit the electromagnetic contribution of the Abrikosov vortex to the free energy since it has a standard form and in the limit $\lambda \gg \xi$, it is independent of  $\rho_0$). In the case $\rho_0>\xi$ we can derive for $F^\text{I(II)}_\text{int}$ explicit expression}
\begin{align}\label{eq:ff_int}
    & F^\text{I(II)}_{\text{int}}\!=\nonumber\!\frac{\!a\!\mp \!g_Nu}{4B_1}\!\!\int\!\!\frac{2J_1\left(1\mp k\right)\left(x^2+y^2\right)}{\left[\left(-1+k\right)x^2-\left(1+k\right)y^2\right]\lambda^2}\\
    & {} \times K^2_1\left(\frac{1}{\lambda}\sqrt{\frac{\left(x-x_0\right)^2}{1+k}+\frac{\left(y-y_0\right)^2}{1-k}}\right)\,dV,
\end{align}
where the upper (lower) sign corresponds to the type I (II) spin vortex, and the integration is performed over the sample volume. {\color{black} The factor before the integral in Eq.~\eqref{eq:ff_int} is equal to the equilibrium value of the GL order parameter in the deformed sample, and the GL coefficients $J_1$ and $k$ reflect the anisotropy of the system.}  The technical details are described in the Appendix \ref{sec:smv}. Thus, $F_\text{int}$ is the part of the free energy associated with the interaction between the vortices. The interaction force is a derivative $\nabla_{\rho_0} F_{\textrm{int}}$. {\color{black} This force decays exponentially when the distance between vortices $\rho_0$ exceeds the London penetration depth $\lambda$ since the modified Bessel function $K_1(x)$ at $x \gg 1$ can be approximated as $K_1(x)\propto e^{-x}/\sqrt{x}$.} 

The physical meaning of the obtained result is as follows. A pattern of the current flowing around the Abrikosov vortex is distorted near the spin vortex's normal core, giving rise to the interaction between vortices. In the case of a usual superconductor with a scalar order parameter, a similar mechanism causes an attraction between a normal inclusion and the vortex~\cite{brandt1995flux}. We show that it is also true for the considered nematic superconductor. 

In general, the interaction between the vortices should include additional contributions of a different nature. The first one arises due to a dependence of the free energy of the Abrikosov vortex on the order parameter $\Delta$, which can vary with coordinates in the strained sample. However, in the considered approximation, $u(r,y)= const$, this term does not contribute to the force $\nabla_{x,y} F_{\textrm{int}}$. In a more general case, it is small if the strain varies over the macroscopic scale $l_u$. The second term is usual short-range pinning on a normal inclusion (with a characteristic scale $\xi$) ~\cite{brandt1995flux}: the Abrikosov and spin vortices have normal cores, and it is thermodynamically favorable to join the cores. Such a term is of significance when $\rho_0<\xi$.

\begin{figure*}[ht]
    \includegraphics[width=\linewidth]{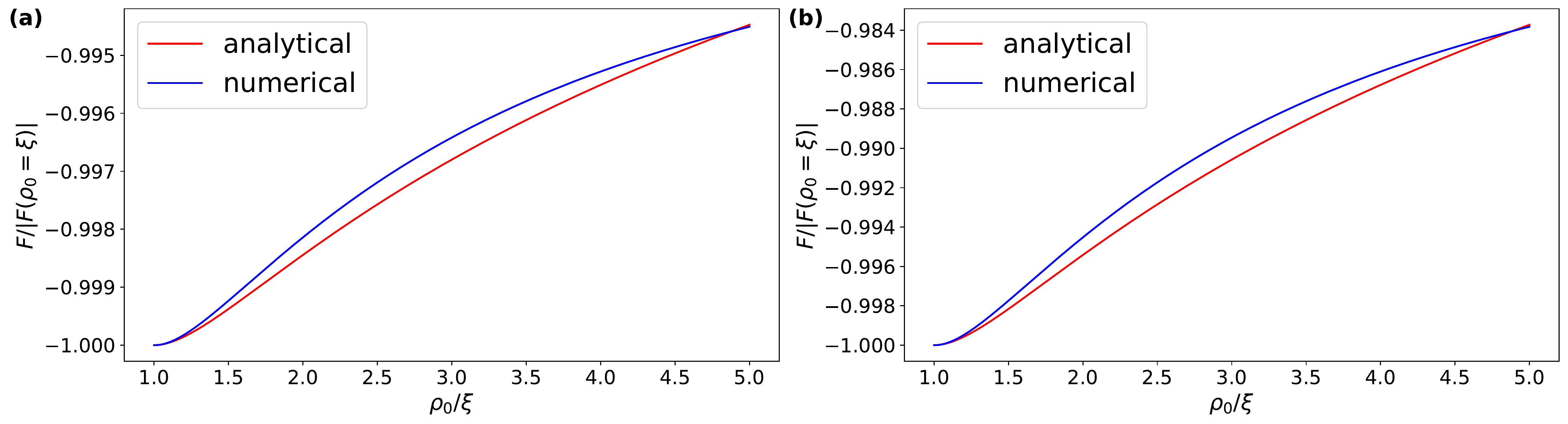}
    \caption{The part of GL free energy responsible for the interaction between the spin and Abrikosov vortices plotted with the following parameters: $a/B_1=-1,\,\,\,\,k =J_1/J_4=0.5,\,\,\,\,\xi/\lambda=0.25,\,\,\,\,\xi/l_u=0.07$. Panel (a) corresponds to the spin vortex of type I at $g_Nu/a=0.25$, and panel (b) corresponds to the spin vortex of type II at $g_Nu/a=-0.25$. The blue line is the calculations by Eq.~\eqref{eq:ff_int}, and the red line is the analytical formula~\eqref{eq:fgl_asymptotic}. 
}
    \label{FIG:2}
\end{figure*}

In Fig.~\ref{FIG:2} we show the function $F^\text{I(II)}_{\text{int}}(\rho_0)$ calculated numerically {\color{black} from Eq.~\eqref{eq:ff_int}. Evidently, the interaction between vortices is significant only when $\rho_0<\lambda$, otherwise, it is exponentially small. When $\xi\leq\rho_0\ll\lambda$ we can derive an analytical formula for $F^\text{I(II)}_{\text{int}}(\rho_0)$ using asymptotic of the modified Bessel functions (see Appendix \ref{sec:smv}):} 
\begin{equation}\label{eq:fgl_asymptotic}
    F^\text{I(II)}_\text{int}=\frac{J_1\pi\left(a\mp g_Nu\right)}{2B_1}\left(\frac{\xi^2}{\rho^2_0}+2\ln{\frac{\rho_0}{\xi}}\right).
\end{equation}
{\color{black} Free energy acquires its minima when $\rho_0=\xi$.}
We see that both in the case of type I and type II vortices, the Abrikosov vortex attracts to their cores if $\rho_0\geq\xi$. When $\rho_0\leq\xi$ the short-range pinning comes into play, and the attraction between the vortices increases significantly~\cite{brandt1995flux}. {\color{black} Qualitative behaviour of the free energy remain the same for all temperatures lower than critical. Temperature only renormilize the length scale $\xi^2(T)\propto 1/(T_c-T)$ and prefactor $a \mp g_Nu\propto T_c-T$}. 

Note, that the interaction of the Abrikosov vortex with the spin vortex has the same nature as usual pinning. However, this interaction is much stronger than the pinning on a point defect. Really, the length $l_z$ of the core of the spin vortex along $z$ direction is equal to the sample size in that direction. This value is much larger than the size $d_z$ of a usual point defect. Accordingly, the force between the vortices is about $l_z/d_z\gg 1$ larger than the force between the Abrikosov vortex and the point defect.       

\section{Analysis of zero energy states}\label{sec:spectrum}

{\color{black}  Localization of the quasi-particles near vortex cores is a common feature of superconductors. In the case of topological superconductivity, of special interest is the existence of zero-energy Majorana states in the spectrum of such quasi-particles (see, e.g., Refs.~\cite{Akzyanov2021,Fu2008,Akzyanov2015}). According to Eq.~\eqref{eq:ff_int}, the spin and mass vortices attract each other. Consequently, the spin vortex, either of type I or II, and the Abrikosov vortex have a common core in the ground state. The order parameter phase in the joint spin-mass vortex differs from the phases of separate spin or mass vortices, see Eqs.~\eqref{eq:order}, which can modify the quasi-particle states near the common core. To clarify this issue, we need a microscopical treatment of the problem.} 

Now, we assume that $l_u\to+\infty$ and seek localized zero-energy solutions of Bogoliubov-de Gennes (BdG) equations near the common vortex core.  {\color{black}Previously, in Ref.\cite{Akzyanov2021}, we have shown that the spin vortex can (or cannot) host Majorana-Kramer's pairs depending on the type of the vortex. The Majorana-Kramer's pairs are protected by time-reversal symmetry. The Abrikosov vortex lifts this symmetry; therefore, the topological properties of the spin-mass vortex would be different.}

The BdG Hamiltonian in the considered case is~\cite{Fu2014} 
\begin{equation}\label{eq:hbdg0}
    H_\text{BdG}(\mathbf{k})=H_0(\mathbf{k})\tau_z+\vec{\Delta}\tau_x,
\end{equation}
where the single-electron Hamiltonian $H_0$ is~\cite{Liu2010}
\begin{equation}
    H_0(\mathbf{k})=-\mu+m\sigma_z+\upsilon\sigma_x\left(s_xk_y-s_yk_x\right)+\upsilon_zk_z\sigma_y.
\end{equation}
Here $\sigma$, $s$, and $\tau$ are the Pauli matrices acting in orbital, spin, and electron-hole spaces, respectively, and the superconducting order parameter is $\vec{\Delta}=\Delta(r)\sigma_y\vec{s}\,\vec{n}$, $\mathbf{k}$ is the momentum, $\mu$ is the chemical potential, $m$ is a single electron gap, and $\upsilon$ and $\upsilon_z$ are the in-plane and transverse Fermi velocities. Following the GL consideration, we choose $\vec{n}$ as $\vec{n}=\left[\cos{\left(\varphi+\nu\pi/2\right)},\sin{\left(\varphi+\nu\pi/2\right)}\right]$, where $\nu=0$ and $\nu=1$ correspond to the type I spin vortex and type II spin vortex, respectively. According to the results of Section~\ref{sec:interaction}, the presence of the Abrikosov vortex in the common core gives rise to an additional phase in the order parameter. Using Eq.~\eqref{eq:order}, we present symbolically this additional phase as $\vec{\Delta}\tau_x\to\vec{\Delta}\tau_xe^{in\theta\tau_z}$.

The spin-mass vortex can be described by introducing a ``defect term'' in the Hamiltonian
\begin{equation}\label{eq:defectterm}
    U^*_\text{SV}U^*_\text{AV}\vec{\Delta}_0U_\text{AV}U_\text{SV},
\end{equation}
where $\vec{\Delta}_0=\Delta(r)\sigma_ys_x\tau_x$ is the superconducting order parameter in the absence of the vortices, $U_\text{SV}=e^{-is_z\left(\varphi/2+\left(\nu-1\right)\pi/4\right)}$ is the spin vortex operator, and $U_\text{AV}=e^{-in\varphi\tau_z/2}$ is the Abrikosov (mass) vortex operator. The spin vortex generates vorticity in the spin space $s$ and can be induced in the Hamiltonian by the transformation $U^*_\text{SV}\Delta_0U_\text{SV}$ \cite{Akzyanov2015,Chiu2016}. The mass vortex generates vorticity in the mass space $\tau$ and can be induced by the transformation $U^*_\text{AV}\Delta_0U_\text{AV}$ \cite{Volovik1999,Chiu2016}. Together, these transformations generate the term associated with the spin-mass vortex.

Further study of possible Majorana states near the vortex core repeats the procedure described in details in Ref.~\cite{Akzyanov2021}. We diagonalize Hamiltonian~\eqref{eq:hbdg0} with order parameter in the form Eq.~\eqref{eq:defectterm}. Then, we perform a set of rather cumbersome transformations and seek zero-energy solutions to the problem. However, we observe that such solutions do not exist either in the case of type I or type II spin vortices. We put all the calculations in Appendix~\ref{sec:e}. Thus, we do not have Majorana fermions in the considered two-vortex system. At the same time, for the spin vortex of type I without the Abrikosov vortex, the Majorana-Kramers pair exists near the vortex core. The obtained results are briefly summarized in Tab.\ref{tab:1}.
\begin{table}[h]
    \caption{The table shows that the existence of the Majorana fermions (MF) in the topological superconductor depends on the topological defects. Here we have four cases: both types of spin vortices (SV) with or without the Abrikosov vortex (AV).}
    \label{tab:1}
    \centering
    \begin{ruledtabular}
    \begin{tabular}{l l l}
            \textrm{SV type} & \textrm{Without AV} & \textrm{With AV}\\
            \colrule
            Type I & 2 MF & No MF \\
            Type II  & No MF & No MF \\
    \end{tabular}
    \end{ruledtabular}

\end{table}
\section{Conclusions}

We analyze the structure of the Abrikosov and strain-induced spin vortices in the nematic superconductor with vector order parameter within the GL approach. We found the conditions under which the nucleation of the spin vortices is possible. We consider the interaction between the spin and mass (Abrikosov) vortices in the nematic superconductor. We show that the Abrikosov vortex attracts to the spin vortex, either of type I or type II. As a result, in the ground state, the mass and spin vortices have a common core. Such a situation is quite similar to the superfluid $^3$He, where the spin-mass vortex was observed \cite{Korhonen1993}. 

The attraction of the vortices reduces the inhomogeneity of the system. The reason for that is simple: in the case of separate vortices, there are two singular points that correspond to the vortex cores where phase gradients diverge. The combined spin-mass vortex has only one core and, therefore, one point where the gradient terms are large. In contrast to the interaction of the Abrikosov vortices, there is no electromagnetic repulsion between spin and mass vortices since the spin vortex does not carry the electrical current. 

We get that the strain should be large enough to generate the spin vortex for a finite radius of the deformed area $l_u$ Eq.~\eqref{eq:crit_strain}. The deformation scale $l_u$ can be either macroscopic $(l_u\gg\xi)$ or microscopic $(l_u\sim\xi)$. However, in the case of small area of deformation $l_u \ll \xi$, even strong deformation does not generate the spin vortex.

{\color{black} The core of the spin-mass vortex can be considered a topological defect in the system. However, no Majorana zero-energy modes are localized near the core in contrast to the case of the spin vortex of type I. Note also that, typically, the Fermi energy is much larger than the value of the order parameter, $E_F=\sqrt{\mu^2-m^2} \gg \Delta$. In this case, as shown for a similar system~\cite{Ziesen2021}, a scale of the minigap between the states localized near the vortex core should be of the order of $\Delta^2/E_F\sim 10^{-3}\Delta$, that is, quite small.}

{\color{black} As we can see from Eq.~\ref{eq:defectterm}, the defect term that introduces the spin-mass vortex in the system can be decomposed as $U_{\textrm{SMV}}=U_{\textrm{HQV}}U_{\textrm{HQV}}$, where $U_{\textrm{HQV}}\propto \exp (i\varphi(s_z+\tau_z)/4)$ corresponds to the half-quantum vortex (HQV)~\cite{Chiu2016}. So, the spin-mass vortex can be formally considered a doubled half-quantum vortex. In Ref.~\cite{How2020}, it was shown that HQVs can appear in a narrow region of parameters in a doped topological insulator. It is an intriguing question whether we can get isolated HQVs in the considered system. However, the main reason why we have a spin vortex in the system is the presence of the strain that forces the nematic superconductor to obey cylindrical symmetry. We assume that the strain is rather strong, and we can apply only the first GL equation to determine the vector structure of the order parameter. Thus, in the considered case here, we can rule out the possibility of forming separate HQVs, while in different geometry, it might be possible.}

The combined spin and Abrikosov vortices can be detected by scanning tunneling microscopy (STM) or scanning SQUID microscopy (SSM). There are two possible scenarios for an experiment. In the first case, we observe the displacement of the Abrikosov vortex lattice under the local force that generates the spin vortex. In the second scenario, we can move the Abrikosov vortex into the strained area $l_u$ using the needle of the STM. Then we again check the displacement of the Abrikosov vortex using the STM or SSM. {\color{black} Control of the nematic superconductivity by the strain was demonstrated in Ref.~\cite{Kostylev2020}. The Abrikosov vortices in doped topological superconductors were observed in STM measurements in Ref.~\cite{Tao2018}. So, we believe that an experimental observation of the spin-mass vortices in doped topological insulators is a feasible task.}




\section*{Acknowledgments}
R.S.A. and A.V.K. acknowledge the support by the Russian Scientific Foundation under Grant No. 22-72-00032.

\appendix

\section{Abrikosov vortex in the nematic superconductor}\label{sec:a}

Here we obtain a solution of the GL equations, which corresponds to an isolated Abrikosov vortex along the $z$ axis located far from the core of the spin vortex and the sample boundaries. In this case, we can assume that the order parameter is constant in the scale of the London penetration depth $\lambda\ll l_u$. If $\vec\Delta$ is independent on the polar angle $\varphi$, we can choose $\vec\Delta=(\Delta,0)$ and the problem is reduced to a standard one for type II superconductors~\cite{Abrikosov1957,tinkham2004introduction}.  
We choose the vector-potential in the form $\mathbf{A}=\left(A_x,A_y,0\right)$. Thus, the GL gradient term Eq.~\eqref{eq: gradient_general} becomes 
\begin{align}\label{eq:avfd1}
    & F_D \!=\nonumber\!\left(\!J_1\!+\!J_4\!\right)\!\hbar^2\!\!\left|\frac{\partial\Delta}{\partial x}\right|^2\!\!\!\!+\!\left(\!J_1\!-\!J_4\!\right)\!\hbar^2\!\left|\frac{\partial\Delta}{\partial y}\right|^2\!\!\!\!+\!i\frac{2e\hbar}{c}A_x\!\left(\!J_1\!+\!J_4\!\right)\\\nonumber
    & {} \times\!\!\left(\!\Delta^*\!\frac{\partial\Delta}{\partial x}\!-\!\Delta\!\frac{\partial\Delta^*}{\partial x}\!\right)\!+\!i\frac{2e\hbar}{c}A_y\!\left(\!J_1\!-\!J_4\!\right)\!\left(\!\Delta^*\!\frac{\partial\Delta}{\partial y}\!-\!\Delta\!\frac{\partial\Delta^*}{\partial y}\!\right)\\
    & {} +\left(\frac{2e}{c}\right)^2\left|\Delta\right|^2\left[A^2_x\left(J_1+J_4\right)+A^2_y\left(J_1-J_4\right)\right].
\end{align}
We make the following transformation 
\begin{equation*}
    \Tilde{x}\!=\!\frac{x}{\sqrt{1\!+\!k}},\!\quad\! \Tilde{y}\!=\!\frac{y}{\sqrt{1\!-\!k}},\!\quad\! A_x\!=\!\frac{\Tilde{A}_x}{\sqrt{1\!+\!k}},\!\quad\! A_y\!=\!\frac{\Tilde{A}_y}{\sqrt{1\!-\!k}},
\end{equation*}
where $k=J_4/J_1<1$. In so doing, we obtain 
\begin{align}\nonumber
    &\frac{F_D}{J_1}\!=\!\hbar^2\!\left(\!\left|\frac{\partial\Delta}{\partial \Tilde{x}}\right|^2\!\!+\!\left|\frac{\partial\Delta}{\partial \Tilde{y}}\right|^2\!\right)\!+\!i\frac{2e\hbar}{c}\!\left[\!\left(\!\Delta^*\!\frac{\partial\Delta}{\partial \Tilde{x}}\!-\!\Delta\!\frac{\partial\Delta^*}{\partial \Tilde{x}}\!\right)\!\Tilde{A}_x\right.\\
    & {} \left.+\!\left(\!\Delta^*\frac{\partial\Delta}{\partial \Tilde{y}}\!-\!\Delta\frac{\partial\Delta^*}{\partial \Tilde{y}}\!\right)\!\Tilde{A}_y\right]\!+\!\left(\frac{2e}{c}\right)^2\!\left(\Tilde{A}^2_x\!+\!\Tilde{A}^2_y\right)\!\left|\Delta\right|^2.
\end{align}
To derive the GL equations, we should minimize the total free energy $F_{GL}=F_0+F_D+F_u+F_M$, Eq.~\eqref{eq: gl_total_1}, with respect to the order parameter and the vector-potential. 

First, we minimize the GL free energy with respect to the complex conjugate order parameter and get
\begin{align}
    \nonumber
    &\frac{\delta_{\Delta^*}F_\text{GL}}{J_1} \!=\!-\hbar^2\!\left(\frac{\partial^2\Delta}{\partial \Tilde{x}^2}+\!\frac{\partial^2\Delta}{\partial \Tilde{y}^2}\right)+i\frac{4e\hbar}{c}\!\left(\!\Tilde{A}_x\!\frac{\partial\Delta}{\partial \Tilde{x}}\!+\!\Tilde{A}_y\!\frac{\partial\Delta}{\partial \Tilde{y}}\!\right)\\\nonumber
    & {} +\left(\frac{2e}{c}\right)^2\left(\Tilde{A}^2_x+\Tilde{A}^2_y\right)\Delta+\frac{a+g_N\left(u_\text{xx}-u_\text{yy}\right)}{J_1}\Delta\\
    & {} +\frac{2B_1}{J_1}\Delta\left|\Delta\right|^2=0.
\end{align}
After a straightforward algebra, we obtain the first GL equation in the form   
\begin{align}\label{eq:gleqD}
    &\nonumber -\nabla^2\Delta+i\frac{4\pi}{\Phi_0}\!\left(\!\tilde{A}_x\frac{\partial\Delta}{\partial\tilde{x}}\!+\!\tilde{A}_y\frac{\partial\Delta}{\partial\tilde{y}}\right)\!+\!\left(\frac{2\pi}{\Phi_0}\right)^2\!\!\left(\tilde{A}^2_x+\tilde{A}^2_y\right)\Delta\\
    & {} +\frac{a+g_N\left(u_\text{xx}-u_\text{yy}\right)}{J_1\hbar^2}\Delta+\frac{2B_1}{J_1\hbar^2}\Delta\left|\Delta\right|^2=0,
\end{align}
where $\Phi_0=\pi\hbar c/e$. In the considered case, the role of the strain reduces to a renormalization of the parameter $a$ in $F_0$.

The second GL equation we derive by minimizing the GL free energy with respect to the vector potential
\begin{align}\label{eq:gleqA}
    &\delta_{\mathbf{\Tilde{A}}}F_{GL}\!=\nonumber\!\frac{1}{4\pi\!\left(1\!-\!k^2\right)}\!\left(\!\nabla\!\!\times\!\!\nabla\!\!\times\!\mathbf{\Tilde{A}}\right)\!
    \\
    & {} \!+iJ_1\!\frac{2e\hbar}{c}
    \left(\Delta^*\nabla\Delta\!-\!\Delta\nabla\Delta^*\right)
    \!+\!2J_1\!\left(\!\frac{2e}{c}\!\right)^2\!\mathbf{\Tilde{A}}\left|\Delta\right|^2\!=\!0.
\end{align}

We write down the GL equations in the coordinates $(\tilde{x},\tilde{y},z)$. It is convenient to rewrite them in the cylindrical coordinates $(\rho,\varphi,z)$ and introduce the modulus and phase of the order parameter $\Delta=|\Delta| e^{i\theta}$ (to be short, we omit below $|...|$). In the new notations we rewrite the GL Eqs.~\eqref{eq:gleqD} and \eqref{eq:gleqA} in the form
\begin{align}\label{eq:gleqpolareta}
\nonumber
   &-\nabla^2\Delta\!+\!\left(\nabla\theta-\frac{2\pi}{\Phi_0}\mathbf{\tilde{A}}\right)^2\!\!\!\Delta
   -\!2i\nabla\Delta\!\left(\!\nabla\theta\!-\!\frac{2\pi}{\Phi_0}\mathbf{\tilde{A}}\!\right)
   \\
   &-i\nabla^2\theta\,\Delta+\frac{a+g_N\left(u_\text{xx}-u_\text{yy}\right)}{J_1\hbar^2}\Delta+\frac{2B_1}{J_1\hbar^2}\Delta^3=0,   
\end{align}
\begin{equation}\label{eq:gleqpolarA}
    \frac{1}{4\pi\hbar^2\left(1-k^2\right)}\left(\nabla\times\nabla\times\mathbf{\tilde{A}}\right)=J_1\frac{4\pi}{\Phi_0}\Delta^2\left(\nabla\theta-\frac{2\pi}{\Phi_0}\mathbf{\tilde{A}}\right).
\end{equation}

The doped topological insulator is a type-II superconductor ~\cite{Yonezawa2019,Kuntsevich2018,KuntsevichAkzyanov2018,Kong2013}. We assume that the GL parameter $\kappa\gg 1$. In this case, the order parameter is a constant if $\rho\gg\xi$ and becomes $\Delta=\Delta_0=-\left[a+g_N\left(u_\text{xx}-u_\text{yy}\right)\right]/2B_1$. Thus, we can calculate the vector potential and the magnetic field from~\eqref{eq:gleqpolarA} following a standard approach for the type-II superconductors~\cite{tinkham2004introduction}. We seek a solution of Eq.~\eqref{eq:gleqpolarA}, which depends only on $\rho$ and place a corresponding delta-function term in the coordinate origin to take into account the vortex core. As a result, we have  from Eq.~\eqref{eq:gleqpolarA}
\begin{equation}
    \vec{H}+\frac{c^2}{32\pi e^2\Delta^2_0J_1\left(1-k^2\right)}\left(\nabla\times\nabla\times\mathbf{H}\right)=\Phi_0\delta(\rho)\vec{e}_z.
\end{equation}
The solution of the latter equation corresponding to the Abrikosov vortex reads
\begin{align}
    H(\rho)=\frac{\Phi_0}{2\pi\lambda^2}K_0&(\rho/\lambda),\quad A(\rho)=-\frac{\Phi_0}{2\pi\lambda}K_1(\rho/\lambda),\\\nonumber
    & {} \quad\frac{1}{\lambda^2}=\frac{32\pi e^2\Delta^2_0J_1\left(1-k^2\right)}{c^2}.
\end{align}
In the chosen gauge $\theta=0$. Near the vortex core, $\rho\leq \xi$, we can readily observe that all the terms related to the magnetic field are of the order of $1/\kappa$ and can be neglected~\cite{tinkham2004introduction}. Then, from the first GL equation~\eqref{eq:gleqpolareta}, we get 
\begin{align}
     f^{\prime\prime}(\Tilde{\rho})+\frac{1}{\Tilde{\rho}}f^{\prime}(\Tilde{\rho})-\frac{1}{\Tilde{\rho}^2}f=f(\Tilde{\rho})^3-f(\Tilde{\rho}),\\\nonumber
     f(\rho)=\frac{\Delta(\rho)}{\Delta_0},\quad\Delta^2_0=-\frac{a\!+\!g_N\left(u_\text{xx}\!-\!u_\text{yy}\right)}{2B_1},\\\nonumber
     \Tilde{\rho}=\frac{\rho}{\xi},\quad\xi^2=\frac{J_1\hbar^2}{2B_1\Delta^2_0}.
\end{align}
Thus, the equation for the order parameter near the core of the Abrikosov vortex is the same as in the case of an s-wave type-II superconductor ~\cite{Abrikosov1957}. Therefore, we have the same asymptotic for order parameter as in the s-wave type-II superconductor and can cut the singularities in $H(\rho)$ and $A(\rho)$ at $\rho\rightarrow 0$ putting $\rho=\xi$.  

\section{GL free energy of the spin vortex}\label{sec:b}

Here we derive the GL free energy of the spin vortex given in Section \ref{sec:sv}.

We assume that the mechanical problem has a cylindrical symmetry, that is, a force produces the strain in the $z$ direction, which acts on a plate sample that has a form of a disc. As stated in Section \ref{sec:sv}, we assume that the strain is constant and $|a|>g_Nu$, and we do not have any magnetic field. In this case, the GL free energy Eq.~\eqref{eq: gl_total_1} consists of the following terms 
\begin{equation*}
    F_0\!=\!a\!\!\left(\!\left|\!\Delta_1\!\right|^2\!+\!\left|\!\Delta_2\!\right|^2\!\right)\!+\!B_1\!\!\left(\!\left|\!\Delta_1\!\right|^2\!+\!\left|\!\Delta_2\!\right|^2\!\right)^2\!\!+\!B_2\!\left|\!\Delta^*_1\Delta_2{-}\Delta_1\Delta^*_2\!\right|^2\!\!\!,
\end{equation*}
\begin{align}
    & F_D\!=\nonumber\!\left(\!J_1\!+\!J_4\!\right)\!\left(\!\frac{\partial\Delta_1}{\partial x}\!\right)^2\!\!\!\!+\!\left(\!J_1\!-\!J_4\!\right)\!\left(\!\frac{\partial\Delta_1}{\partial y}\!\right)^2\!\!\!\!+\!\left(\!J_1\!-\!J_4\!\right)\!\left(\!\frac{\partial\Delta_2}{\partial x}\!\right)^2\!\!\!\!\\\nonumber
    & {} +\!\left(\!J_1\!+\!J_4\!\right)\!\left(\!\frac{\partial\Delta_2}{\partial y}\!\right)^2+2J_4\!\left(\!\frac{\partial\Delta_1}{\partial x}\frac{\partial\Delta_2}{\partial y}\!+\!\frac{\partial\Delta_1}{\partial y}\frac{\partial\Delta_2}{\partial x}\!\right),
\end{align}
\begin{align*}
    F_\text{u}=\nonumber&g_Nu(r,z)\left(\left|\Delta_1\right|^2-\left|\Delta_2\right|^2\right)\cos{2\varphi}\\
    & {} + g_Nu(r,z)\left(\Delta^*_1\Delta_2+\Delta_1\Delta^*_2\right)\sin{2\varphi}.
\end{align*}
The spin vortex order parameter can be written as \cite{Akzyanov2021}
\begin{equation}
    \!\!\!\!\!\!\!\vec{\Delta}_\text{I}\!\!=\!\!\Delta(r,z)\!\left(\cos{\varphi},\sin{\varphi}\right),
\,\,\vec{\Delta}_\text{II}\!\!=\!\!\Delta(r,z)\!\left(-\sin{\varphi},\cos{\varphi}\right),
\end{equation}
We substitute these order parameters into the free energy of the spin vortex. Thus, to get the energy of the spin vortex and the critical strain, we have to evaluate the following integral over the sample volume
\begin{equation*}
F_\text{SV}=\int  dV\left(F_0+F_D+F_u\right).    
\end{equation*}
We put $\Delta=\Delta_0$ and in the cylindrical coordinates with the accuracy $\xi^2/l_u^2\ll 1$ obtain 
\begin{align}
    & F^\text{I(II)}_\text{SV}=\nonumber\int\limits_{\xi_\text{I(II)}}^{l_u}\int\limits_{0}^{2\pi} \frac{1}{4B_1r}\left[\mp 2a\left(\pm J_1+J_4+g_Nr^2u\right)\right.\\
    & {} \left.+g_Nu\left(\pm 2J_1+2J_4+3g_Nr^2u\right)\right]\,dr\,d\varphi,
\end{align}
where the upper (lower) sign corresponds to the type I (II)  spin vortex, $\xi_\text{I(II)}=\xi\sqrt{1\pm k}$ is the coherence length for the type I(II) spin vortex.   
Thus, we get the GL spin vortex free energy in the form 
\begin{align}
    F^\text{I(II)}_\text{SV}=&\nonumber\frac{\pi}{4B_1}\left[g_Nu\left(\mp2a+3g_Nu\right)l^2_u\right.\\
    & {} \left.-4\left(J_1\pm J_4\right)\left(a\mp g_Nu\right)\ln{\frac{l_u}{\xi_\text{I(II)}}}\right].
\end{align}

\section{Attraction between the spin and mass vortices}\label{sec:smv}

Here we evaluate the GL interaction energy between the Abrikosov and spin vortices Eq.~\eqref{eq:ff_int}: 
\begin{align}\label{eq:fint_c}
    & F^\text{I(II)}_{\text{int}}\!=\!\frac{\!a\!\mp \!g_Nu}{4B_1}\!\!\int\!\!\frac{2J_1\left(1\mp k\right)\left(x^2+y^2\right)}{\left[\left(-1+k\right)x^2-\left(1+k\right)y^2\right]\lambda^2}\\\nonumber
    & {} \times K^2_1\left(\frac{1}{\lambda}\sqrt{\frac{\left(x-x_0\right)^2}{1+k}+\frac{\left(y-y_0\right)^2}{1-k}}\right)\,dV,
\end{align}
where $(x_0,y_0)$ are the coordinates of the Abrikosov vortex. To be short, we present the calculation details for the case of type I spin vortex. For the type II vortex, the derivation is similar. 

We introduce the polar coordinates $(\rho,\theta)$ related to $(x,y)$ by the following transformation 
\begin{align}
    & x=\left(\rho\cos{\theta}+\rho_0\cos{\theta_0}\right)\sqrt{1+k},\\\nonumber
    & {} y=\left(\rho\sin{\theta}+\rho_0\sin{\theta_0}\right)\sqrt{1-k},\\\nonumber
    & {} x_0=\rho_0\cos{\theta_0},\\\nonumber
    & {} y_0=\rho_0\sin{\theta_0}.
\end{align}
Then, the integral \eqref{eq:fint_c} becomes
\begin{equation}
    \!\!\!F^\text{I}_\text{int}\!=\!-\frac{2J_1\!\left(\!a\!-\!g_Nu\!\right)}{4B_1}\!\!\!\!
    \int\limits_{\xi}^{l_u}\!\!\!\rho K^2_1(\!\rho/\lambda\!)\sqrt{\!1\!-\!k^2}d\rho\!\!\int\limits_{0}^{2\pi}\!\!\!R(\rho,\theta,\rho_0,\theta_0)\,d\theta,
\end{equation}
where
\begin{align}\label{eq:fint_complex}
    & \nonumber\int\limits_{0}^{2\pi}\!\!\!R(\rho,\theta,\rho_0,\theta_0)\!\,d\theta\!=\!\!\!\int\limits_{0}^{2\pi}\!\!\left\{\!\frac{\rho^2+\rho^2_0+k\rho^2\cos{2\theta}}{\left(\!1\!+\!k\!\right)\!\lambda^2\!\left(\rho^2\!+\!\rho^2_0\!+\!2\rho\rho_0\!
    \cos{\![\theta\!-\!\theta_0]}\right)}\right.\\
    & {} \left.+\frac{\rho_0\left[k\rho_0\cos{2\theta}+2\rho\left\{\cos{(\theta-\theta_0)}+k\cos{(\theta+\theta_0)}\right\}\right]}{\left(1+k\right)\lambda^2\left(\rho^2+\rho^2_0+2\rho\rho_0\cos{[\theta-\theta_0]}\right)}\right\}\!\,d\theta.
\end{align}
To evaluate the latter integral, we turn to integration in the complex plane $z$ using the following substitution
\begin{equation}
    z = e^{i\theta},\quad d\theta = \frac{dz}{iz},\quad z_0=e^{i\theta_0.}
\end{equation}
The integral~\eqref{eq:fint_complex} becomes
\begin{align}\label{eq:r_int_complex}
    & \nonumber\int\limits_{0}^{2\pi}R(\rho,\theta,\rho_0,\theta_0)\,d\theta\\\nonumber
    & {} =\!-i\!\!\!\!\oint\limits_{\left|z\right|=1}\!\!\!\!k\!\left\{\frac{\rho^2\!\left(z^4\!+\!1\right)\!z^2_0\!+\!\rho^2_0z^2\!\left(z^4_0\!+\!1\right)\!+\!2\rho\rho_0\!\left(z^2z^2_0+1\right)}{2\left(1+k\right)\lambda^2z^2z_0\left(\rho_0z+\rho z_0\right)\left(\rho z+\rho_0z_0\right)}\right.\\
    & {} \left.+\frac{1}{z\lambda^2\left(1+k\right)}\right\}\,dz
\end{align}
The function under the integral has three poles 
\begin{equation}
    z_1=0,\quad z_2=-\frac{z_0\rho}{\rho_0},\quad z_3=-\frac{z_0\rho_0}{\rho}.
\end{equation}
We can put $z_0=1$ since the choice of position $\theta=0$ is arbitrary due to rotational symmetry of the problem. We calculate the integral~\eqref{eq:r_int_complex} using the residue theorem. If $\rho_0/\rho\leq1$ we get 
\begin{align}
    \int\limits_{0}^{2\pi}R(\rho,\theta,\rho_0,\theta_0)\,d\theta=2\pi i\left[\underset{z=z_1}{\text{Res}} R+\underset{z=z_3}{\text{Res}} R\right]=\frac{2\pi}{\left(1+k\right)\lambda^2}
\end{align}
and if $\rho_0/\rho>1$ we have
\begin{align}
    &\nonumber\int\limits_{0}^{2\pi}R(\rho,\theta,\rho_0,\theta_0)\,d\theta=2\pi i\left[\underset{z=z_1}{\text{Res}} R+\underset{z=z_3}{\text{Res}} R\right]\\
    & {} =\frac{2\pi\left[\rho^2_0-k\left(\rho^2-\rho^2_0\right)\right]}{\left(1+k\right)\lambda^2\rho^2_0}.
\end{align}
We consider the case $\rho_0\geq\xi$. Then, the integral \eqref{eq:fint_c} becomes
\begin{align}
    & \nonumber F^\text{I}_\text{int}=F^\text{I}_\text{int}(\rho_0\geq\xi)\\\nonumber
    & {} =-\frac{J_1\pi\left(a-g_Nu\right)\sqrt{1-k^2}}{B_1\lambda^2\left(1+k\right)}\left[\int\limits_{\rho_0}^{l_u}\rho K^2_1(\rho/\lambda)\,d\rho\right.\\
    & {} \left.+\int\limits_{\xi}^{\rho_0}\rho\left[1+k\left(1-\rho^2/\rho^2_0\right)\right]K^2_1(\rho/\lambda)\,d\rho\right].
\end{align}
In the case of interest $l_u< \lambda$, when the vortices interact with each other, we can use the asymptotic $K_1(\rho/\lambda)\simeq\lambda/\rho$ and derive 
\begin{equation}
     F^\text{I(II)}_\text{int}=\frac{J_1\pi\left(a\mp g_Nu\right)}{2B_1}\left(\frac{\xi^2}{\rho^2_0}+2\ln{\frac{\rho_0}{\xi}}\right).
\end{equation}
Here we add the result for the type II spin vortex, which can be derived using a similar approach.

\section{Bogoliubov-de Gennes equations}\label{sec:e}

Here we present a detailed analysis of the possible existence of the zero-energy states near the core of the spin-mass vortex. We can consider only the states with $k_z=0$ and rewrite the Hamiltonian Eq.~\eqref{eq:hbdg0} with ``defect term'' Eq.~\eqref{eq:defectterm} in the polar coordinates: 
\begin{align}\label{eq:hbdg1}
    & H_\text{BdG}=\nonumber-\mu\tau_z+m\sigma_z\tau_z+i\upsilon\sigma_xs_x\tau_z\left[e^{i\left(\varphi+\pi/2\right)s_z}\nabla_r\right.\\
    & {} \left.-\frac{1}{r}e^{i\varphi s_z}\nabla_{\varphi}\right]+\Delta\sigma_ys_x\tau_xe^{i\left[\left(s_z+n\tau_z\right)\varphi+\nu\pi/2s_z\right]}.
\end{align}
The dependence of the Hamiltonian on the polar angle $\varphi$ can be removed by the transformation 
\begin{equation}\label{eq:tansform_hbdg1}
    \psi(r,\varphi)=\exp{\left[i\left(l-s_z/2-n\tau_z/2\right)\varphi\right]}\frac{\tilde{\psi}(r)}{\sqrt{r}}, 
\end{equation}
where $l$ is an orbital number and $\psi$ is the eight-component spinor $\left(f_{1\uparrow},f_{1\downarrow},f_{2\uparrow},f_{2\downarrow},h_{1\uparrow},h_{1\downarrow},h_{2\uparrow},h_{2\downarrow}\right)^\text{T}$. Here, 1 and 2 are the orbital indices, $\uparrow$ and $\downarrow$ are the spin projections, and $f$ and $h$ represent the electron and hole states. The wave function is a single valued if $l=0,\pm1,\pm2,...$\,. Thus, the Hamiltonian \eqref{eq:hbdg1} becomes 
\begin{align}
    & H_\text{eff}=\nonumber e^{-i\left(l-s_z/2-n\tau_z/2\right)\varphi}H_\text{BdG}e^{i\left(l-s_z/2-n\tau_z/2\right)\varphi}\\\nonumber
    & {} =\!-\!\mu\tau_z\!+\!m\sigma_z\tau_z\!+\!i\upsilon\sigma_xs_x\tau_z\!\left[is_z\nabla_r\!-\!\frac{i}{r}\!\left(l\!-\!s_z/2\!-\!n\tau_z/2\right)\!\right]\\
    & {} +\Delta\sigma_ys_x\tau_xe^{i\frac{\nu\pi}{2}s_z}.
\end{align}
The Hamiltonian Eq.~\eqref{eq:hbdg0} has a symmetry $\left[H,\sigma_zs_z\right]=0$. Thus, there is a basis in which the Hamiltonian is decomposed in two spin-orbital blocks with $\sigma_zs_z\hat{\Psi}_{\pm}=\pm\hat{\Psi}_{\pm}$ and in this basis the operator $\sigma_zs_z$ is diagonal. Here $\hat{\Psi}_{+}=\left(\Psi_{+},0\right)^\text{T}$, and $\hat{\Psi}_{-}=\left(0,\Psi_{-}\right)^\text{T}$ where 
\begin{equation*}
    \Psi_{+(-)}=\left(h_{1(2)\downarrow},h_{1(2)\uparrow},f_{2(1)\downarrow},f_{2(1)\uparrow}\right)^\text{T}.
\end{equation*}
The transformation that decomposes the Hamiltonian $H_\text{eff}$ can be presented in the matrix form as
\begin{equation}
P=
    \begin{pmatrix}
    0 & 0 & 0 & 0 & 0 & 0 & 0 & 1 \\
    0 & 0 & 0 & 0 & 0 & 0 & 1 & 0 \\
    0 & 0 & 0 & 1 & 0 & 0 & 0 & 0 \\
    0 & 0 & 1 & 0 & 0 & 0 & 0 & 0 \\
    0 & 1 & 0 & 0 & 0 & 0 & 0 & 0 \\
    1 & 0 & 0 & 0 & 0 & 0 & 0 & 0 \\
    0 & 0 & 0 & 0 & 0 & 1 & 0 & 0 \\
    0 & 0 & 0 & 0 & 1 & 0 & 0 & 0
    \end{pmatrix}.
\end{equation}
After transformation, $P^\dag H_\text{eff}P$, we obtain the Hamiltonian $H_\text{eff}$ in the block-diagonal form 
\begin{equation}
    H_\text{eff}=
    \begin{pmatrix}
    H_{\beta=+1} & 0 \\
    0 & H_{\beta=-1}
    \end{pmatrix},
\end{equation}
where 0 corresponds to the $4\times4$ zero matrix and
\begin{equation}\label{eq:hrho}
    H_{\beta}=\left(\mu+\beta mc_z+i\upsilon c_y\nabla_r\right)\tau_z-\frac{n\upsilon}{2r}c_x+\beta\Delta c_y\tau_xe^{-i\frac{\pi\nu}{2}c_z}, 
\end{equation}
where $\beta=\pm 1$ and we assume that $l=0$ since we are only interested in the zero-energy states. Here $c_i$ is the Pauli matrix that acts in the spin-orbital space, $\tau_i$ acts in particle-hole space, the case $\nu=0$ corresponds to the type I spin vortex, and $\nu=1$ corresponds to the type II spin vortex. 

To simplify the further analysis, we can assume that $m\rightarrow 0$. In this case, we have a topologically equivalent system. If such a system has no zero-energy modes, therefore, the system does not have these modes when $m\neq 0$.   

In the case of the type I spin vortex and Abrikosov vortex ($\nu=0$), the Hamiltonian anticommutes with the operator $c_y\tau_y$, $\left\{H,c_y\tau_y\right\}=0$. The operator $c_y\tau_y$ is diagonalized by the transformation
\begin{equation}
    P_0=\frac{1}{2}
    \begin{pmatrix}
    1 & 0 & -1 & 0\\
    0 & -1 & 0 & 1\\
    0 & 1 & 0 & 1\\
    1 & 0 & 1 & 0
    \end{pmatrix}.
\end{equation}
We apply this transformation to the Hamiltonian $H_\beta$ \eqref{eq:hrho}, $P^\dag_0H_{\beta}P_0$, and obtain
\begin{equation}
    H^\text{I}_{\beta}=
    \begin{pmatrix}
    0 & H^\text{I}_{t=-1}\\
    H^\text{I}_{t=+1} & 0
    \end{pmatrix},
\end{equation}
where
\begin{equation}\label{eq:ht_nu_0}
    H^\text{I}_t=-\mu\kappa_z+t\left(i\beta\Delta\kappa_z+\frac{in\upsilon}{2r}\kappa_y-\upsilon\kappa_x\nabla_r\right).
\end{equation}
Here $t=\pm1$ and $\kappa_i$ are the Pauli matrices that act in the space $\vec{L}_{\nu=0}=\left(L_1,L_2\right)^\text{T}=\left(f_{1(2)\uparrow}\pm h_{2(1)\downarrow},f_{1(2)\downarrow}\mp h_{2(1)\uparrow}\right)^\text{T}$ for $\beta=+1(-1)$. Thus, for the type I spin vortex and the Abrikosov vortex, we have four blocks of $2\times2$ equations. 

In the case of the type II spin vortex and Abrikosov vortex ($\nu=1$), the Hamiltonian anticommutes with the operator $c_y\tau_x$, $\left\{H,c_y\tau_x\right\}=0$. The operator $c_y\tau_x$ is diagonalized by the transformation
\begin{equation}
    P_1=\frac{1}{2}
    \begin{pmatrix}
    i & 0 & -i & 0\\
    0 & i & 0 & -i\\
    0 & 1 & 0 & 1\\
    1 & 0 & 1 & 0.
    \end{pmatrix}
\end{equation}
Then we apply this transformation to the Hamiltonian $H_\beta$ \eqref{eq:hrho}, $P^\dag_1H_{\beta}P_1$, and obtain
\begin{equation}
    H^\text{II}_{\beta}=
    \begin{pmatrix}
    0 & H^\text{II}_{t=-1}\\
    H^\text{II}_{t=+1} & 0
    \end{pmatrix},
\end{equation}
where
\begin{equation}\label{eq:ht_nu_1}
    H^\text{II}_t=-\mu\kappa_z+t\left(-i\beta\Delta+\upsilon\kappa_y\nabla_r+\frac{in\upsilon}{2r}\kappa_x\right).
\end{equation}
Here $t=\pm1$ and $\kappa_i$ are the Pauli matrices that act in the space $\vec{L}_{\nu=1}=\left(L_1,L_2\right)^\text{T}=\left(f_{1(2)\uparrow}\mp ih_{2(1)\downarrow},f_{1(2)\downarrow}\mp ih_{2(1)\uparrow}\right)^{T}$ for $\beta=+1(-1)$. Thus, for the type II spin vortex and the Abrikosov vortex, we also have four blocks of $2\times2$ equations. 

Now we seek zero-energy solutions to the BdG equations. Thus, we solve equations $H_{t}\vec{L}=0$. We assume that the superconducting order parameter is a step function, namely $\Delta=0$ for $r\leq\xi$ and $\Delta\ne0$ for $r>\xi$. For the type I spin vortex ($\nu=0$) and one Abrikosov vortex ($n=1$), we have 
\begin{align}
    & \left(L^{\prime\prime}_2+\frac{1}{r}L^{\prime}_2\right)\upsilon^2+\left(\mu-it\beta\Delta\right)^2L_2=0,\\ \nonumber
    & {} L_1\left(\mu-it\beta\Delta\right)=-t\upsilon L^{\prime}_2.
\end{align}
Solutions regular at $r\leq\xi$ can be presented in the form 
\begin{align}\label{eq:r_l_xi}
    &L_1=C_1J_1\left(\frac{rt\mu}{\upsilon}\right),\\\nonumber
    & {} L_2=C_1J_0\left(\frac{rt\mu}{\upsilon}\right),
\end{align}
where $J_m(x)$ are the $m$-th order Bessel functions. Solutions regular at $r>\xi$ are
\begin{align}\label{eq:r_g_xi1}
    &\text{for}\quad\beta=+1:\\\nonumber
    & {} L_1\!=\!\tilde{C}_1\!\left[iJ_1\!\left(\!\frac{rt\left(\mu\!-\!it\Delta\right)}{\upsilon}\!\right)\!+\!Y_1\!\left(\!\frac{rt\left(\mu\!-\!it\Delta\right)}{\upsilon}\!\right)\!\right],\\\nonumber
    & {} L_2\!=\!\tilde{C}_1\!\left[iJ_0\!\left(\!\frac{rt\left(\mu\!-\!it\Delta\right)}{\upsilon}\!\right)\!+\!Y_0\!\left(\!\frac{rt\left(\mu\!-\!it\Delta\right)}{\upsilon}\!\right)\!\right].
\end{align}
\begin{align}\label{eq:r_g_xi2}
    &\text{for}\quad\beta=-1:\\\nonumber
    & {} L_1\!=\!\tilde{C}_1\!\left[\!-iJ_1\!\left(\!\frac{rt\left(\mu\!+\!it\Delta\right)}{\upsilon}\!\right)\!+\!Y_1\!\left(\!\frac{rt\left(\mu\!+\!it\Delta\right)}{\upsilon}\!\right)\!\right],\\\nonumber
    & {} L_2\!=\!\tilde{C}_1\!\left[\!-iJ_0\!\left(\!\frac{rt\left(\mu\!+\!it\Delta\right)}{\upsilon}\!\right)\!+\!Y_0\!\left(\!\frac{rt\left(\mu\!+\!it\Delta\right)}{\upsilon}\!\right)\!\right].
\end{align}
Here $Y_m(x)$ are the $m$-th order Neumann functions. Solutions \eqref{eq:r_l_xi} and \eqref{eq:r_g_xi1}-\eqref{eq:r_g_xi2} cannot be matched at $r=\xi$. Thus, there are no zero-energy solutions localized near the spin-mass vortex core.

Now we consider the type II spin vortex ($\nu=1$) and one Abrikosov vortex ($n=1$). For this case, the transformed BdG equations read 
\begin{align}
    & \left(L^{\prime\prime}_2+\frac{1}{r}L^{\prime}_2\right)\upsilon^2+\left[\mu^2+\Delta^2\right]L_2=0,\\ \nonumber
    & L_1\left(\mu+it\Delta\beta\right)=-it\upsilon L^{\prime}_2.
\end{align}
The solutions of the latter equation are 
\begin{align}
    & L_1=\\\nonumber
    & {} \!\frac{i\mu\!+\!t\Delta\beta}{\sqrt{\mu^2\!+\!\Delta^2}}\!\left[\!C_2J_{1}\!\left(\!\frac{rt\sqrt{\mu^2\!+\!\Delta^2}}{\upsilon}\!\right)\!+\!C_3Y_{1}\!\left(\!\frac{rt\sqrt{\mu^2\!+\!\Delta^2}}{\upsilon}\!\right)\!\right],\\\nonumber
    & {} L_2\!=\!C_2J_{0}\!\left(\!\frac{rt\sqrt{\mu^2\!+\!\Delta^2}}{\upsilon}\!\right)\!+\!C_3Y_{0}\!\left(\!\frac{rt\sqrt{\mu^2\!+\!\Delta^2}}{\upsilon}\!\right).
\end{align}
These solutions are not regular at $r>\xi$. Thus, there are no zero-energy solutions localized near the spin-mass vortex core.

\bibliography{ref}

\end{document}